\documentclass{aa} 
\usepackage{graphicx} 
\usepackage{times}
\usepackage{natbib} 
\bibpunct{(}{)}{;}{a}{}{,} 

\newcommand{\msuns}{$\mathrm{{M}_\odot}$ }
\newcommand{\mjup}{$\mathrm{{M}_J}$}
\newcommand{\mjups}{$\mathrm{{M}_J}$ }
\newcommand{\mearth}{$\mathrm{{M}_\oplus}$}

\begin{document}  
  \title{Dust flow in gas disks in the presence of embedded planets} 
  \author{Sijme-Jan Paardekooper \inst{1} \and Garrelt Mellema \inst{2,1}} 
  \offprints{S. J. Paardekooper\\\email{paardeko@strw.leidenuniv.nl}} 
  \institute{Leiden Observatory, Postbus 9513, NL-2300 RA Leiden, 
             The Netherlands\\
  \email{paardeko@strw.leidenuniv.nl} \\ \and
             ASTRON, Postbus 2, NL-7990 AA Dwingeloo, The Netherlands\\
  \email{gmellema@astron.nl}} 
  \date{Draft Version \today}
  
  \abstract{}{We study the dynamics of gas and dust in a protoplanetary disk
    in the presence of embedded planets. We investigate the conditions for 
    dust-gap formation in terms of particle size and planetary mass. We also
    monitor the amount of dust that is accreted by the planet relative to the 
    amount of gas, which is an important parameter in determining the 
    enrichment of solids in giant planets compared to the solid content of 
    the central star.}
    {We use a new two-fluid hydrodynamics code to solve the flow equations for 
    both gas and dust. For the gas, we use a Godunov-type scheme with an 
    approximate Riemann solver (the Roe solver). The dust is treated as a 
    pressureless fluid by essentially the same numerical method as is used for 
    the gas.}
    {We find that it only takes a planet of $0.05$ Jupiter masses to open up a 
    gap in a disk with a significant population of mm-sized particles. Dust 
    particles larger than 150 $\mathrm{\mu m}$ participate in gap formation. 
    We also find that the formation of the gap severely slows down dust 
    accretion compared to that in the gas. Therefore, it is not possible to 
    enrich a newly formed giant planet in solids, if these solids are contained
    in particles with sizes from 150 $\mathrm{\mu m}$ to approximately 10 cm.}
    {}
     
  
  \maketitle
  
\section{Introduction}
Dust plays a major role in planet formation, theoretically as well as 
observationally. First of all, terrestial planets consist mainly of solid 
material. They are giant agglomerates of interstellar dust particles, and 
therefore the behaviour of these small particles is important during the first 
stages of planet formation. Whether they grow to kilometer-sized bodies by slow
coagulation or through the gravitational instability of a thin dust layer in 
the midplane of the disk \citep{1973ApJ...183.1051G} is still an open question.
A mixture of equal amounts of gas and dust is subject to streaming 
instabilities \citep{2005ApJ...620..459Y}, and it is not clear whether the 
resulting turbulent structure will prohibit gravitational instabilities or 
enhance them \citep{2006ApJ...636.1121J}. To answer this question, 
multi-component simulations are needed in which gas and dust are able to 
evolve separately.

Secondly, giant planets can be formed either by gas accretion onto a solid 
core \citep{1996Icar..124...62P} or by direct gravitational fragmentation of 
the disk \citep{1997Sci...276.1836B}. In the core accretion model, dust growth 
is of critical importance to build up a massive core before the gas disk 
dissipates. The gravitational instability model does not rely on the existence
of such a core, but still, accretion of dust in a later stage will determine 
the enrichment in solids with respect to the solar nebula, which is an
important observational constraint \citep{2004jpsm.book...35G}.

Finally, dust is the major source of continuum radiation in the disk. 
Therefore, infra-red and sub-millimeter emissions mainly indicate the density
and temperature of the dust disk. The relation of the dust to the density of 
the gas disk can be quite complicated if the disk has a non-trivial structure, 
especially if the dust particles are only marginally coupled to the gas.

It is therefore very important to study dust dynamics in gas disks.
Axisymmetric disks can be handled analytically, and various authors have 
studied dust migration in these disks \citep{1977MNRAS.180...57W,
2002ApJ...581.1344T}. Particles move both inward and outward under the 
influence of gas drag, depending on their distance from the midplane, but the 
average migration is inward \citep{2002ApJ...581.1344T}.
\cite{2003ApJ...598.1301H} studied dust migration in a non-uniform 
axisymmetric nebula, and \cite{2001ApJ...557..990T} investigated the effect
of a sharp disk edge on dust migration. The overall conclusion is that the 
structure of the gas disk can have a significant effect on the dust 
distribution, and that the distribution of gas and dust can differ 
substantially.

When the gas distribution becomes more complicated, one has to rely on 
numerical simulations. \cite{2001ApJ...551..461S} studied dust evolution during
the initial collapse of the nebula with a two-dimensional multi-fluid 
hydrodynamics code to obtain the large-scale dust distribution. 
\cite{2005ApJ...634.1353J} studied the effect of MHD turbulence on the dust 
distribution. These multi-fluid calculations are computationally very 
expensive, requiring more than twice the computing time for ordinary 
gas-dynamical simulations, even if one solves for only a single dust particle 
size. 

To compare dust observations with gas-dynamical simulations, but at
the same time avoid the large computational effort of these multi-fluid 
calculations, one can assume a fixed dust-to-gas ratio and solve only for the 
gas dynamics only. The effects of an embedded planet on a gaseous disk have 
been studied numerically before by multiple authors 
\citep[e.g.][]{1999MNRAS.303..696K,2000MNRAS.318...18N,1999ApJ...526.1001L,
2002A&A...385..647D}. \cite{2002ApJ...566L..97W} used these results 
to simulate observations of gaps in protoplanetary disks. Their calculations 
are valid only for the smallest particles that couple extremely well to the 
gas.

However, in a protoplanetary disk where grain growth is very important
\citep[see, for example,][]{2005A&A...434..971D}, a significant fraction of 
the dust population is large enough to be only marginally coupled to the gas. 
\cite{2004A&A...425L...9P} showed that including the full gas and dust dynamics
leads to a dramatic evolution of the dust component in the disk. Planets not 
massive enough to open up a gap in the gas disk do so in the dust disk, which 
should make the orbit of even $0.05$ \mjups planets visible to ALMA (the 
Atacama Large Millimeter Array).

In this paper we continue our investigation of dust flow in gas disks around
embedded planets, with more attention to detail than in 
\cite{2004A&A...425L...9P}. In Sect. \ref{secequations}, we review the 
equations governing dust flow and gas-dust interaction. Section \ref{secmethod}
is devoted to the numerical method, and we discuss the initial and boundary 
conditions in Sect. \ref{secinitial}. In Sect. \ref{secres}, we present the 
results. Section \ref{secDisc} is reserved for a short discussion, and we 
conclude in Sect. \ref{seccon}.

\section{Basic equations}
\label{secequations}
Protoplanetary disks are fairly thin, i.e. the vertical thickness $H$
is small compared with the distance $r$ from the centre of the disk.
Typically we use $H/r = h = 0.05$. It is therefore convenient to average
the equations of motion vertically and to work with vertically averaged
quantities only. 

The governing equations are then solved in a cylindrical coordinate frame
$(r,\phi)$, centred on the central star and corotating with the embedded
planet that has a Keplerian angular velocity $\Omega$. 

\subsection{Gas}
The equations that control the gas evolution are described in detail 
in \cite{paardekooper03}. We do not solve the energy equation, but 
instead use an isothermal equation of state:
\begin{equation}
P=c_\mathrm{s}^2 \Sigma_\mathrm{g},
\end{equation}
where $\Sigma_\mathrm{g}$ is the surface density of the gas and $P$ is
the vertically averaged pressure.

The isothermal sound speed $c_\mathrm{s}$ is, in hydrostatic equilibrium, 
directly related to the disk thickness:
\begin{equation}
\label{eqsound}
c_\mathrm{s} = h~v_\mathrm{K},
\end{equation}
where $v_\mathrm{K}$ is the Keplerian velocity.

We use a constant kinematic viscosity $\nu$, which is set by assuming an
$\alpha$-parameter \citep{1973A&A....24..337S} of $0.004$ at the location
of the planet:
\begin{equation}
\nu = \left. \alpha c_\mathrm{s} H \right |_{\mathrm{r=1}}.
\end{equation}

\subsection{Dust}
We treat the dust as a pressureless fluid; its evolution is governed
by conservation of mass and, in absence of external sources, conservation
of radial and angular momentum. These can be written in the following 
compact form:
\begin{equation}
\label{eqEuler}
\frac{\partial \vec{W}}{\partial t} + \frac{\partial \vec{F}}{\partial
r}+ \frac{\partial \vec{G}}{\partial \phi} = \vec{S},
\end{equation}
where $\vec{W}$ is called the state vector and where $\vec{F}$ and $\vec{G}$
are the fluxes in the radial and azimuthal direction, respectively.
$\vec{S}$ is called the source term. 

The components of the state, fluxes, and source-term vectors can be 
written in the following form:
\begin{eqnarray}
\vec{W}=r(\Sigma,~\Sigma v_r,~\Sigma v_\phi) \\ 
\vec{F}=r(\Sigma v_r,~\Sigma v_r^2,~ \Sigma v_r v_\phi)\\ 
\vec{G}=r(\Sigma v_\phi,~ \Sigma v_r v_\phi,~ \Sigma v_\phi^2)
\end{eqnarray}
\begin{eqnarray}
\vec{S}=\left( \begin{array}{c} 0 \\
\label{eqfull}
\Sigma r^2 (v_\phi+\Omega)^2 - 
\Sigma r \frac{\partial{\Phi}}{\partial{r}} + r f_{\mathrm{d},r}\\
 -2 \Sigma v_r (\Omega+v_\phi) - \frac{\Sigma}{r}
\frac{\partial{\Phi}}{\partial{\phi}} + r f_{\mathrm{d},\phi}
 \end{array}\right).
\label{eq2D}
\end{eqnarray}
Here, $\Sigma$ denotes the dust surface density, $v_r$ and $v_\phi$ the radial
and angular velocity, respectively, $\Omega$ the angular velocity of the
coordinate frame (corotating with the planet), and $\Phi$ the gravitational
potential of the central star and the planet. The gravitational potential
of the planet is softened over a fraction $0.2$ of the Roche lobe of the 
planet. The potential also contains terms due to the non-inertial nature of 
the coordinate system. The drag forces are incorporated through 
$f_{\mathrm{d},r}$ and $f_{\mathrm{d},\phi}$.

The major difference between the equations for the gas and the dust is the
absence of pressure in the latter. In this sense, the dust fluid behaves
like a gas that is always moving with supersonic velocity. This implies that
near shock waves, where the gas goes from sonic to supersonic flow and where
large density and velocity gradients are present, dust and gas behave in very 
different ways. 

The basic building block for Godunov-type hydrodynamic solvers is the Riemann 
problem, consisting of two stationary states separated by a discontinuity. If
we consider the two states as belonging to neighbouring grid cells, with the
discontinuity at the interface between the cells, the analytic solution of 
the Riemann problem can be used to define an interface flux between the cells. 
Popular schemes of this type are the Piecewise Parabolic Method \citep[PPM,][]
{1984JCoPh..54..115W} and the Roe solver \citep{1981...............}.

\begin{figure}
\resizebox{\hsize}{!}{\includegraphics[bb=35 10 285 250]{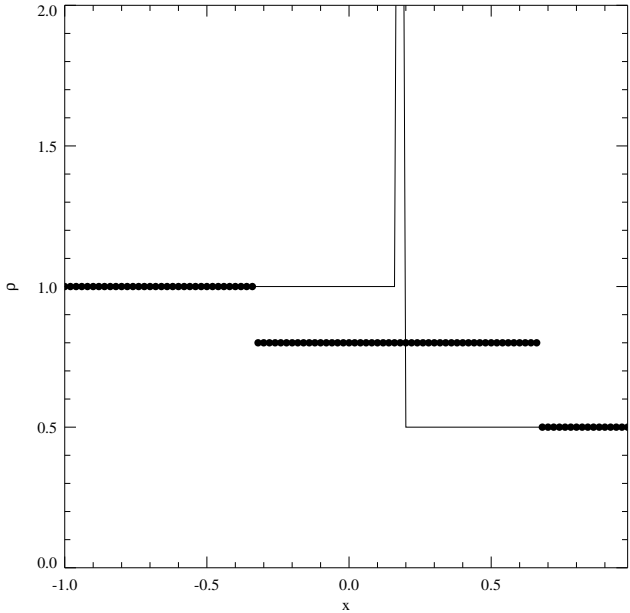}}
\caption{Solution to the Riemann problem with initial conditions $\rho_L=1.0$,
$\rho_R=0.5$, $u_L=0.5$, and $u_R=0.25$. The solid line represents the dust
density and the dots represent the gas density, both at $t=0.5$. For the gas,
the speed of sound was set to $1.0$.}
\label{fig1}
\end{figure}

For pressureless fluids, the solution to the Riemann problem differs 
significantly from the case with pressure. Figure \ref{fig1} shows the 
analytic solutions for the gas (dots) as well as the dust density (solid line),
with initial conditions $\rho_L=1.0$, $\rho_R=0.5$, $u_L=0.5$, and $u_R=0.25$. 
The gas solution consists of two waves moving at speeds $\hat u -c_\mathrm{s}$
and $\hat u + c_\mathrm{s}$, where
\begin{equation}
\label{eqshockspeed}
\hat u=\frac{\sqrt{\rho_L}~u_L+\sqrt{\rho_R}~u_R}{\sqrt{\rho_L}+\sqrt{\rho_R}}.
\end{equation}
If we now let
$c_\mathrm{s}\rightarrow 0$ to obtain a pressureless fluid, the separation
between the waves becomes smaller and smaller, until they overlap at
$c_\mathrm{s}=0$. The overlapping waves create a delta-function singularity
that moves with velocity $\hat u$ \citep[e.g.][]{leveque}. The singularity 
arises because Eq. \ref{eqEuler} is only {\it weakly} hyperbolic, in 
contrast with the gas hydrodynamical equations, which are {\it strictly} 
hyperbolic.

Figure \ref{fig1} deals with the case $u_L > u_R$. When $u_L < u_R$, the 
solution consists of the two states moving away from each other with a vacuum
in between. Note that for both cases, the flux is well defined everywhere
but at the position of the delta function. 

The solution to the Riemann problem not only forms the basis for our numerical 
method (see Sec. \ref{secmethod}), but from Fig. \ref{fig1} it is also clear 
that gas and dust react very differently to \emph{physical} shocks, which may
lead to gas-dust separation when shock waves are present in the computational 
domain. 

\subsection{Gas-dust interaction}
\label{subsecgasdust}
The interaction between the gas fluid and the dust fluid occurs only through
the drag forces. The nature of the drag force depends on the size of the
particles with respect to the mean free path of the gas molecules. We
consider only spherical particles with radius $s$. Because
the main constituent of the gas is molecular hydrogen, for the mean free path
we can write:
\begin{equation}
\lambda=\frac{m_{\mathrm{H_2}}}{\pi \rho_\mathrm{g} {r_\mathrm{H_2}}^2}.
\end{equation}
When the particle size is small, compared to $\lambda$, we are in the Epstein 
regime, whereas $s \gg \lambda$ corresponds to the Stokes regime. For 
particles that move subsonically through the gas, the transition occurs 
around $\lambda / s = 4/9$. For typical midplane densities, particles larger 
than approximately $50$ cm move in the Stokes regime. In this paper we will 
focus on particles smaller than $\sim 1$ cm, and therefore we can safely 
use the Epstein drag law.

In the Epstein regime, when the relative velocity of gas and dust is much 
smaller than the local sound speed, the drag force is written as:
\begin{equation}
\vec{f}_{\mathrm{d}}=
-\Sigma \frac{\Omega_\mathrm{K}}{T_\mathrm{s}} \Delta \vec{v}, 
\end{equation}
where $\Omega_\mathrm{K}$ is the Keplerian angular velocity, $\Delta \vec{v}$
is the velocity difference between gas and dust, and $T_\mathrm{s}$
is the dimensionless stopping time \citep{2002ApJ...581.1344T}:
\begin{equation}
T_\mathrm{s}=\sqrt{\frac{\pi}{8}} \frac{\rho_\mathrm{p} s 
v_\mathrm{K}}{\rho_\mathrm{g} r c_\mathrm{s}}.
\end{equation}
Here, $\rho_\mathrm{p}$ is the particle internal density, $s$ is the radius
of the particle, and $c_\mathrm{s}$ is the sound speed. We have used 
$\rho_\mathrm{p}=1.25$ $\mathrm{g~cm^{-3}}$, and the isothermal sound speed
is given by Eq. \ref{eqsound}. The gas density $\rho_\mathrm{g}$ is found
by using $\rho_\mathrm{g}=\Sigma_\mathrm{g}/2 H$.

For particles that move very supersonically through the gas, the drag force 
becomes:
\begin{equation}
\vec{f}_{\mathrm{d}}=
-\Sigma \frac{3}{4} \sqrt{\frac{\pi}{8}}\frac{\Omega_\mathrm{K}}{T_\mathrm{s}} 
\frac{|\Delta \vec{v}|}{c_\mathrm{s}} \Delta \vec{v}. 
\end{equation}
A standard way to connect the supersonic to the subsonic regime is 
\citep{1975ApJ...198..583K}:
\begin{equation}
\label{eqdrag}
\vec{f}_{\mathrm{d}}=
-\Sigma \frac{\Omega_\mathrm{K}}{T_\mathrm{s}}
\sqrt{1+\frac{9\pi}{128} \frac{|\Delta \vec{v}|^2}{c_\mathrm{s}^2}} 
\Delta \vec{v}. 
\end{equation}
See \cite{2003A&A...399..297W} for the general case.

Only very massive planets of approximately 1 \mjups are able to accelerate
the larger particles to velocities that are comparable to the speed of sound. 
For the low-mass planets considered in \cite{2004A&A...425L...9P}, the subsonic
version of the drag force proved to be sufficient.

Particles with $T_\mathrm{s} \ll 1$ are well-coupled to the gas, and the radial
drift velocity can be written as \citep{2002ApJ...581.1344T}:
\begin{equation}
\label{eqvdrift}
v_{r,\mathrm{drift}}=-\eta T_\mathrm{s} v_\mathrm{K},
\end{equation}
where $\eta$ is the ratio of the gas pressure gradient to the stellar 
gravity in the radial direction:
\begin{equation}
\label{eqeta}
\eta = -h^2 \frac{\partial \log P}{\partial \log r} = 
h^2\left( 1-\frac{\partial \log \Sigma_\mathrm{g}}{\partial \log r}\right).
\end{equation}
The direction in which the dust moves with respect to the gas is determined
by the sign of $\eta$, and in general the pressure gradient is negative, 
making $\eta$ positive, which leads to an inward migration of the dust 
particles. However, if there is a large positive density gradient 
($\partial \log \Sigma_\mathrm{g} / \partial \log r > 1$, see Eq. \ref{eqeta}),
as there is near the outer edge of a density gap, the dust particles can move 
outward. On the other hand, if there is an extra large negative density 
gradient, as is the case near the inner edge of a density gap, the dust moves 
inward with higher velocity. 

Thus, we expect that near the gap edges, the larger particles decouple from 
the gas, altering the dust-to-gas ratio in that region. Furthermore, the
drag forces may not be large enough to prevent gas-dust separation by
shocks. Both processes should play a role in the dust dynamics near the
embedded planet.

\begin{figure*}
\includegraphics[bb=5 10 455 315,width=17cm]{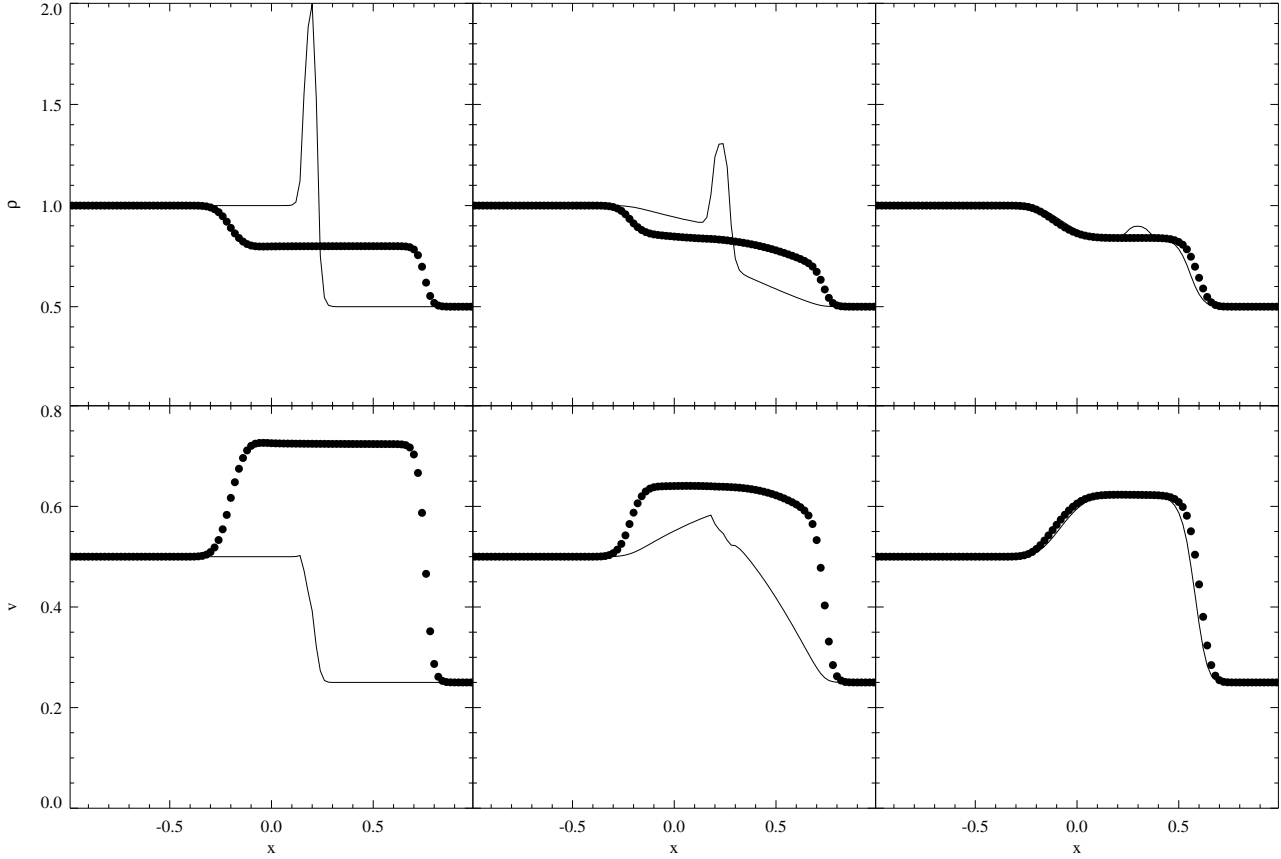}
\caption{Numerical results for the Riemann problem of Fig. \ref{fig1} for gas 
(dots) and dust (solid line). The top row shows the density, the 
bottom row the velocity. Left panels: no gas-dust interaction. Middle panels:
$T_s=0.5$. Right panels: $T_s=0.05$.}
\label{fig2}
\end{figure*}

\section{Numerical method}
\label{secmethod}
\subsection{Gas and dust advection}
For the evolution of the gas component, we use the RODEO method described by
\cite{paardekooper03}. It is a second-order Eulerian hydrodynamics
code, which uses an approximate Riemann solver \citep{1981...............}.
It derives from a general relativistic method \citep{1995A&AS..110..587E},
and it is especially suitable to treat non-Cartesian, non-inertial
coordinate systems. A module for Adaptive Mesh Refinement (AMR) can be used
to obtain high resolution near the planet.

For the dust, we use the method outlined in \cite{leveque}, 
which has the advantage of being similar to the method used for the gas.
Below, we only consider the radial direction. The azimuthal direction is 
handled in the same way. 

For every cell interface, we consider the Riemann problem defined by a left
state $\vec W_\mathrm{L}$ and a right state $\vec W_\mathrm{R}$ with 
corresponding fluxes.
When $v_{\mathrm{r,L}}<0<v_{\mathrm{r,R}}$, the solution to the Riemann 
problem gives us a vacuum state at the cell interface, so we set the 
interface flux $\vec F_{\mathrm{int}}=\vec 0$. Otherwise, we compute the
shock speed $\hat v$ from Eq. \ref{eqshockspeed}, and compute the interface
flux according to
\begin{equation}
\label{eqflux}
\vec F_\mathrm{int}=\left\{ \begin{array}{ll}
       \vec F_\mathrm{L} & \mbox{if $\hat v > 0$};\\
       (\vec F_\mathrm{L}+\vec F_\mathrm{R})/2 ~~~~~& \mbox{if $\hat v = 0$};\\
       \vec F_\mathrm{R} & \mbox{if $\hat v < 0$}. \end{array} \right.
\end{equation}
In the first and the last case, the cell interface is respectively left and
right of the delta shock, allowing for an easy definition of the interface
flux. When $\hat v=0$, however, the delta shock is exactly at the cell 
interface. We then choose to distribute this shock equally over both cells.

To put this in the language of \cite{paardekooper03}, we write 
$\vec{F}=\mathcal{A} \vec{W}$, where the matrix $\mathcal{A}$ is given by:
\begin{equation}
\mathcal{A}=\left( \begin{array}{ccc} v_r & 0 & 0 \\ 
                                      0 & v_r & 0 \\ 
                                      0 & 0 & v_r
\end{array} \right).
\end{equation}
The eigenvalues of this matrix are:
\begin{equation}
\lambda_1 = \lambda_2 =\lambda_3 = v_r.
\end{equation}
The corresponding eigenvectors read:
\begin{eqnarray}
\vec{e}_1 & = & \left(1,0,0 \right) \nonumber \\ 
\vec{e}_2 & = & \left(0,1,0 \right) \\ 
\vec{e}_3 & = & \left(0,0,1 \right).~~~~~~~~~ \nonumber
\end{eqnarray}
A vector $\vec{\Delta}\equiv(\Delta_\rho,\Delta_r,\Delta_\phi)$ can be
projected on these eigenvectors using the following projection
coefficients, found by solving the system
$\mathcal{C}\vec{b}=\vec{\Delta}$, where $\mathcal{C}$ is the matrix
with the eigenvectors:
\begin{eqnarray}
b_1 &=& \Delta_\rho \nonumber\\ 
b_2 &=& \Delta_r  \\ 
b_3 &=& \Delta_\phi \nonumber.
\end{eqnarray}
To approximate the matrix $\mathcal{A}$ at an interface of two grid cells,
we use Roe-averages. If we then project the flux difference across the
interface we are considering onto the eigenvectors of $\mathcal{A}$:
\begin{equation}
\vec{F}_{\rm{R}}-\vec{F}_{\rm{L}}=\sum b_{\rm{k}} \vec{e}_{\rm{k}},
\end{equation}
we can then define the first order interface flux just as it was defined in 
the gas case:
\begin{equation}
\label{eqfirstorder}
\vec F_{\mathrm{int}} =
\frac{1}{2}(\vec{F}_{\rm{L}}+\vec{F}_{\rm{R}})-
\frac{1}{2}\sum \sigma_{\rm{k}} b_{\rm{k}}
\vec{e}_{\rm{k}},
\end{equation}
where $\sigma_{\rm{k}}=\rm{sign}(\lambda_k)$. When $\lambda_{\rm{k}}=0$, we 
take $\sigma_{\rm{k}}=0$. It is easy to show that this
version of the interface flux corresponds exactly with Eq. \ref{eqflux}.
Having the interface flux in this form, we can immediately write down a 
second-order interface flux that is again analogous to the gas case:
\begin{equation}
\label{eqsecorder}
\vec{F}_{\mathrm{int}} =
\frac{1}{2}(\vec{F}_{\rm{L}}+\vec{F}_{\rm{R}})-
\frac{1}{2}\sum (\sigma_{\rm{k}} a_{\rm{k}}
-(\sigma_{\rm{k}}-\nu_{\rm{k}})\psi_{\rm{k}})\lambda_{\rm{k}} \vec{e}_{\rm{k}},
\end{equation}
where $\nu_{\rm{k}} = \lambda_{\rm{k}} \Delta t/\Delta r$ and $\psi$ is
the flux limiter. We have used the same limiter as in \cite{paardekooper03} 
for both gas and dust in all simulations.

\subsection{Accretion}
We model accretion onto the planet by taking mass away from the grid at the 
location of the planet. There are two parameters governing the amount of
mass taken away at each time-step: the accretion area and the accretion rate.
We take the accretion area to be a circle with a radius equal to a tenth of
the Roche lobe of the planet, and the accretion rate to be such that the time 
scale for emptying this region is equal to three fifths of the orbital time 
scale of the planet. The same parameters were used in 
\cite{2002A&A...385..647D} and \cite{paardekooper03}. 
 
\subsection{Source terms}
All the source terms except the drag forces are integrated using stationary 
extrapolation \citep{paardekooper03}. The combination of exact 
extrapolation for the angular momentum and approximate extrapolation for the
other geometrical source terms was shown to give reliable results for 
single-fluid calculations of the planet-disk problem. Because these results
were not affected by increasing the resolution, we can conclude that using
this treatment of the source terms is accurate enough. However, when other 
forces besides gravity are included, it is not guaranteed that a stationary 
solution exists. \cite{2001A&A...371..205S} found that the source terms 
arising in two-fluid calculations are best integrated analytically. Therefore, 
the drag forces are incorporated separately using an ordinary differential 
equation:
\begin{equation}
\frac{d}{dt}(\Sigma \vec{v}) = \vec{f}_{\mathrm{d}}.
\end{equation}
During this step, we keep the dust and gas density fixed. Using Eq. 
\ref{eqdrag}, we can write an ordinary differential equation for the velocity 
difference between gas and dust:
\begin{equation}
\frac{d}{dt}(\Delta \vec{v}) = -\alpha \sqrt{1+\beta \Delta \vec{v}^2} 
\Delta \vec{v},
\end{equation}
where $\alpha=\frac{\Omega_{\mathrm{K}}}{T_\mathrm{s}}$ and 
$\beta=\frac{9\pi}{128 c_\mathrm{s}^2}$. This equation can be integrated 
analytically to give
\begin{equation}
\Delta \vec{v}(t)=\frac{2 \Delta \vec{v}_0 \exp{(-\alpha t)}}
{1+\sqrt{1+\beta \Delta \vec{v}_0^2}-\frac{\beta \vec{v}_0^2\exp{(-2\alpha t)}}
{1+\sqrt{1+\beta \Delta \vec{v}_0^2}}}.
\end{equation}
Note that when $\beta=0$, we recover the ordinary exponential decay of the 
velocity difference. Using conservation of total momentum:
\begin{equation}
\Sigma \vec v + \Sigma_g \vec v_g = \mathrm{constant},
\end{equation}
we can solve for the evolution of the velocities of gas and dust separately,
while keeping the total momentum conserved exactly.

\subsection{Test problem}
Gasdynamical codes are often tested against a simple Riemann problem 
\citep{1978JCoPh..27....1S}, also known as the shock tube. In this section, we 
test the method for dust advection against the Riemann problem shown in Fig. 
\ref{fig1}, using a 100-zone grid. We also study the effect of the interaction 
between gas and dust. The initial dust-to-gas mass ratio is 1. The 
simulations were run until $t=0.5$, which took approximately $50$ time-steps.

In Fig. \ref{fig2}, we show the results for three different stopping times:
$T_\mathrm{s}=\infty$ (left panels), $T_\mathrm{s}=0.5$ (middle panels), and 
$T_\mathrm{s}=0.05$ (right panels). We see that for the case of no interaction,
we reproduce the result of Fig. \ref{fig1} for gas and dust. The delta shock is
smeared over approximately $4$ computational cells, but it moves at the correct
speed. Note that because the velocity is less than the sound speed everywhere,
the waves in the gas are not shock waves, but are simply sound waves.

For the middle panels, the stopping time is comparable to the sound-crossing
time and the dust velocity increases towards the gas velocity, 
while by conservation of momentum, the gas velocity decreases with respect to 
the left panels. The velocities approach each other the fastest at the position
of the delta shock which has decreased in amplitude because the gas and dust 
densities also approach each other.

When we decrease the stopping time by a factor of ten, the densities and 
velocities of gas and dust are almost equal everywhere. In the right panels
of Fig. \ref{fig2}, there is only a small bump left of the delta shock, and
the gas has lost enough momentum to the dust to visibly slow down the sound
waves. Note that this only happens because the gas and dust densities were 
equal initially. For an interstellar dust-to-gas ratio of $1:100$, the gas is 
not affected by the dust. 

\subsection{Limits} 
\label{secLim}
The approach presented here cannot be used in all circumstances. To begin with,
the continuum form for the equations of motion is only valid if two 
requirements are met. First, there need to be enough particles in the volume 
under consideration (in our case, one grid cell) to work with 
average states. Second, there need to be enough interparticle collisions to 
make such an average state meaningful. For a dust fluid, both criteria are 
only fullfilled if the particles are well-coupled to the gas 
\citep{2004ApJ...603..292G}. In other words, we can only work with particles 
for which $T_\mathrm{s} \ll 1$ everywhere in the computational domain. For 
larger dust boulders, a particle-based method is more appropriate
\citep{2005astro.ph.12272J,2005MNRAS.364L..81F}.

On the other hand, when the stopping time is very small, the source terms
due to gas-dust friction become stiff. Even though we integrate these
source terms in an implicit way, which makes the method \emph{stable} for
these small particles, it does not necessarily give us the correct answer. 
This is because the gas source terms due to pressure are dealt with in the 
advection step, and pressure is precisely what causes gas-dust separation. 
Therefore, even for $T_\mathrm{s}=0$, the dust would numerically slowly drift 
inward, while according to Eq. \ref{eqvdrift}, its radial velocity should be 
zero. The range 
in $T_\mathrm{s}$ that we can consider is therefore also bounded from below.
The time scale for the advection (the hydrodynamical time-step,
determined by the Courant-Friedrichs-Lewy (CFL) condition) should not be much 
larger than the time scale for gas-dust coupling. Combining the two limits, 
we can conclude that we can only consider particles for which the stopping 
time satisfies:
\begin{equation}
\Omega_\mathrm{K} \Delta t \approx T_\mathrm{s} \ll 1.
\label{eqTlimit}
\end{equation}
In practice, for the gas densities considered here (see Sect. 
\ref{secinitial}),
Eq. \ref{eqTlimit} limits the particle size from below at approximately 
$150~\mathrm{\mu m}$ and from above at approximately $1$ cm. In all our 
simulations we checked that lowering the time-step $\Delta t$ did not 
influence the results.

It is important to realize that Eq. \ref{eqTlimit} should hold true everywhere
on the grid. When the gas density changes significantly, for example due
to the formation of a gap, the stopping time can vary by orders of 
magnitude. The condition $T_\mathrm{s} \ll 1$ for gap-opening planets makes 
the simulations very expensive. To do the calculations in a reasonable amount 
of time, we restrict our parameter space to non-gap-opening planets. Note, 
however, that gap-opening planets also pose problems for particle-based methods
because very many particles are needed to achieve enough resolution inside the 
gap.

The two-dimensional nature of these simulations also puts conditions on the 
size of the parameter space we can consider. When the Roche lobe of the planet 
is smaller than the disk thickness, three-dimensional effects come into play, 
lowering accretion and migration rates for example, 
\citep{2003ApJ...586..540D}. However, 
global three-dimensional multi-fluid calculations are beyond current 
computational resources. Furthermore, because of dust settling, the dust disk 
will be much thinner than the gas disk, depending on the particle size. 
The two-dimensional approach will therefore cause fewer discrepancies in the 
dust distribution. Still, one has to be very careful when interpreting results 
for planets with $M < 0.1$ \mjup.

\section{Initial and boundary conditions}
\label{secinitial}
The standard numerical resolution we use is $(n_r, n_\phi) = $ $(128,384)$.
This way, the cells close to the planet are of equal size in the radial
and in the azimuthal directions. We do not resolve the Roche lobe using
a low resolution like this, but this resolution is sufficient to capture
the long-term global evolution of the disk. For the runs with dust accretion, 
we used 4 levels of AMR to achieve a resolution that is 16 times higher near 
the planet. 

As mentioned before, we use an aspect ratio of $h=0.05$, and
the initial gas surface density distribution is constant, with 
$\Sigma_\mathrm{g}=34$ g 
$\mathrm{cm^{-2}}$. This surface density is appropriate for a disk 
with $0.01$ \msuns within $100$ AU at approximately $7$ AU, between the 
location of Jupiter and Saturn in the solar system, or at 13 AU in the Minimum
Mass Solar Nebula. Note, however, that the fundamental parameter is the 
stopping time, and our results can therefore be scaled to disks of arbitrary 
mass by adopting a different grain size for a given $T_\mathrm{s}$. 

The initial dust-to-gas ratio is $0.01$. Our distance 
unit is the radius of the planet's orbit, and in this unit the inner boundary 
is at $r=0.4$ and the outer boundary at $r=2.5$. 

It is easy to see that for a constant surface density the stopping time
is also constant:
\begin{equation}
\label{eqstopinit}
T_\mathrm{s}=0.0461~\left(\frac{s}{\mathrm{cm}}\right),
\end{equation}
where the size of the particles $s$ is given in cm. 

The gas rotates with a slightly sub-Keplerian velocity due to the radial 
pressure gradient (note that even when there is no density gradient, there is 
a pressure gradient caused by the gradient in the sound speed). The dust 
rotates exactly with the Keplerian velocity, initially. All radial 
velocities are taken to be zero.

The constant kinematic viscosity is set so that $\alpha=0.004$ at $r=1$ in all 
models. For the disk described above, the critical planet mass for gap opening 
is about $1$ Jupiter mass ($1$ \mjup) \citep{1999ApJ...514..344B}.
We varied the mass of the planet between 0.001 \mjups and 1.0 \mjup.

The boundary conditions are the same as in \cite{paardekooper03}, deriving
from \cite{1996MNRAS.282.1107G}. They are specifically designed to be 
non-reflecting, so that all waves generated in the simulated region leave
the computational domain without influencing the interior. 

\begin{figure}
\resizebox{\hsize}{!}{\includegraphics[bb=0 10 285 250]{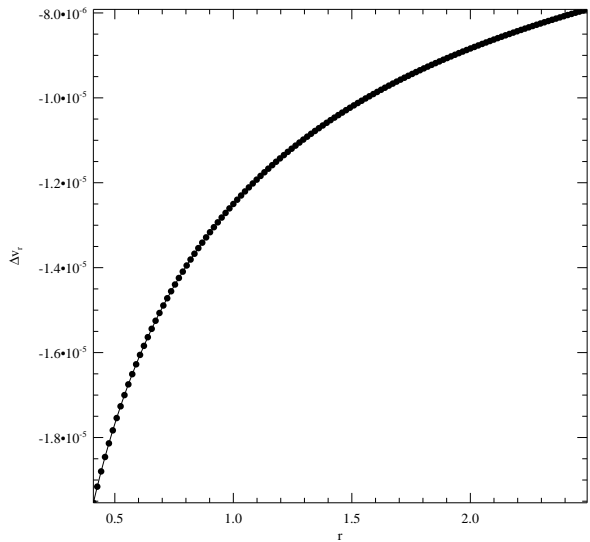}}
\caption{Relative radial velocity of gas and dust (1 mm) in a disk 
without a planet. The dots represent the numerical solution, and the solid line
represents Eq. \ref{eqvdrift}}.
\label{fig3}
\end{figure}
\begin{figure}
\resizebox{\hsize}{!}{\includegraphics[]{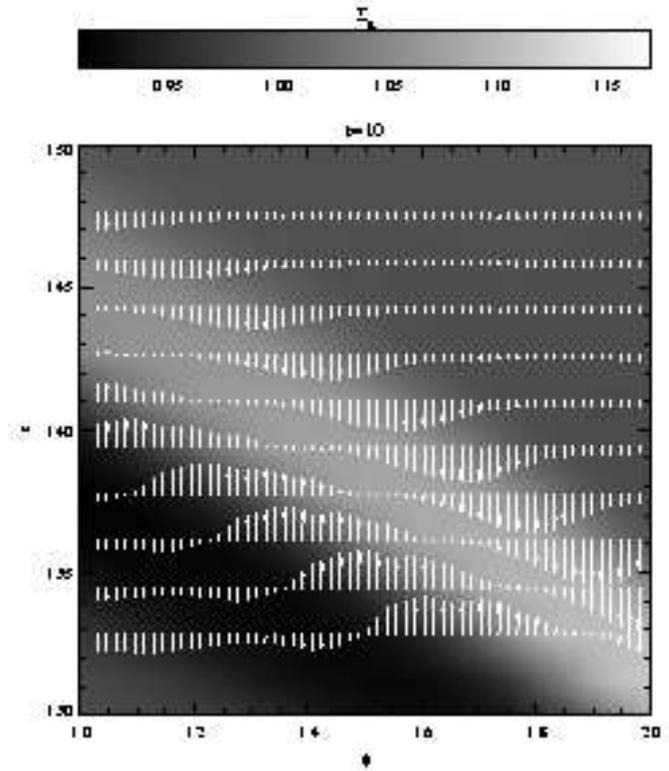}}
\caption{Close-up on a spiral density wave, with relative velocity arrows
superimposed. Outside of the spiral wave, the dust drift is directed inward,
according to Eq. \ref{eqvdrift}. Close to the spiral, the particles move
towards the centre of the wave. For 5 mm dust particles, as shown here,
the maximum relative velocity is $2.6~10^{-6}$, in units of the orbital 
velocity of the planet.}
\label{fig4}
\end{figure}
\begin{figure}
\resizebox{\hsize}{!}{\includegraphics[bb=45 10 285 255]{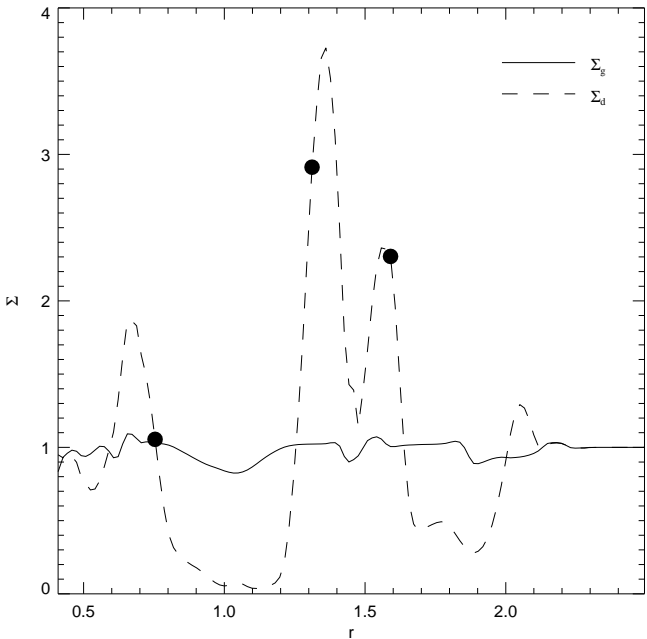}}
\caption{Surface density of gas and dust ($s=0.1$ cm) after 500 orbits of
a $0.1$ \mjups planet at $\phi=0$ (opposite to the planet). The dust 
density is multiplied by 100. The dots indicate the positions of the 2:3, 
the 3:2, and the 2:1 resonances.}
\label{fig5}
\end{figure}

\section{Results}
\label{secres}
\subsection{Axisymmetric disk}
To begin with, we consider a disk without a planet. The analytic solution for
the drift velocity is given in Eq. \ref{eqvdrift}. Figure \ref{fig3} shows
this analytic solution for particles of 1 mm, as well as the numerical 
solution after $10$ orbits of the gas at $r=1$. The analytical solution is 
indistinguishable from the numerical result, showing that the method 
handles the large source terms arising due to the small $T_\mathrm{s}$ very
well. The angular velocity of the dust is always equal to the angular
velocity of the gas. The Coriolis force starts to play a role only when the 
relative radial velocity becomes very large.  

\subsection{Dust response to a spiral wave}
An embedded planet perturbs the gas disk by launching spiral density waves. 
In this section, we investigate the dust response to such a single wave. 
In Fig. \ref{fig4}, we show a close-up on part of the spiral pattern generated
by the planet. The velocity arrows indicate the relative velocity of gas
and dust. In the upper right direction, the density and velocity of the gas 
are close to their initial values, and in this region, the dust drifts slowly
inward, according to Eq. \ref{eqvdrift}. In the lower left part of the figure
the dust particles drift inward again, but at a higher speed, due to the lower 
gas density. At the location of the spiral wave, gas density and velocity are
such that the dust always moves towards the centre of the wave. This is not
surprising in view of Eq. \ref{eqeta}, which states that in an axisymmetric
disk, dust particles will always move towards the highest pressure. Note that
even in this perturbed disk, the angular velocities of gas and dust are still 
almost equal: the particles move only radially through the gas. Therefore,
the dust response to a spiral wave is the same as the response to an 
axisymmetric density wave, but stretched in the radial direction. The effect
of a density wave, as shown in Fig. \ref{fig4}, remains very local and is of
minor importance for the global disk evolution.

\subsection{Global disk evolution}
\label{secGlob}
\cite{2004A&A...425L...9P} showed that planets of $\sim 0.1$ \mjups are 
able to decouple the evolution of the gas disk and the dust disk. More
specifically, they showed that the dust particles tend to move away from the 
orbit of the planet, creating a deep annular gap. Furthermore, the final dust 
disk appeared to be structured near mean motion resonances. In this 
section, we further explain the mechanism for structuring the disk.

First of all, Fig. \ref{fig5} shows a radial density cut through a disk
perturbed by a $0.1$ \mjups planet after 500 orbits. Again, as in 
\cite{2004A&A...425L...9P}, we see a clear gap and a density enhancement 
at the position of the 2:1 mean motion resonance.

Note that since $T_\mathrm{s} \ll 1$, the dust particles are never influenced
by the planet directly at the radii of the indicated resonances, they only 
react to the gas density and velocity. The mechanism for structuring the dust 
disk is therefore intrinsically different from ordinary resonance capture, 
which is the usual explanation for the existence of the Plutinos in the Kuiper
belt \citep{2002ARA&A..40...63L}.

\begin{figure}
\includegraphics[bb=40 10 285 245]{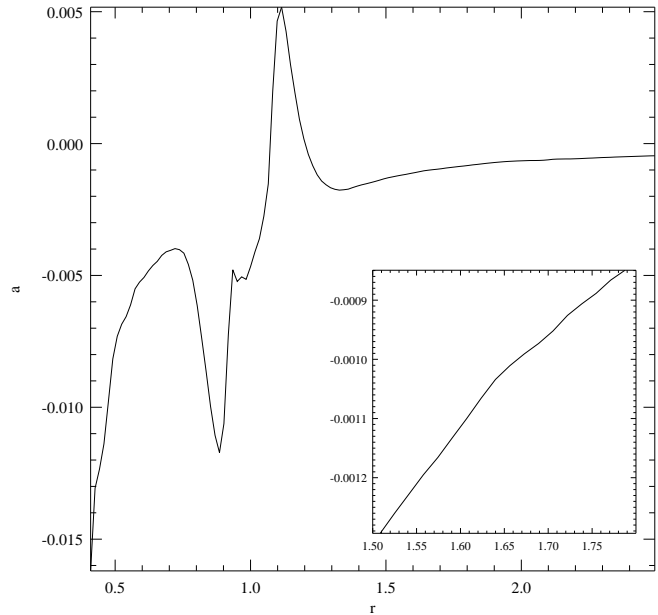}
\caption{Azimuthally averaged radial acceleration $a$ of dust particles after 
100 orbits of a $0.1$ \mjups planet. The close-up shows $a$ in the outer 
region of the disk. The 2:1 mean motion resonance is visible
as a small bump in $a$.}
\label{fig6}
\end{figure}

In Fig. \ref{fig6}, we plot the radial acceleration of dust particles, 
excluding the drag force:
\begin{equation}
a=r (v_\phi+\Omega)^2 - \frac{\partial{\Phi}}{\partial{r}}.
\end{equation}
Figure \ref{fig6} shows the azimuthally averaged $a$, with a close-up on the 
outer region of the disk. The first thing to note is that the 
dust is always accelerated inward, except for the region around $r=1.1$. 
Inside the planet's orbit, the acceleration is very negative, indicating that 
the dust will quickly move toward the inner boundary. 

The large inward and outward accelerations near $r=0.9$ and $r=1.1$, 
respectively, are due to the inner and outer edges of the density dip carved 
in the gas. The pressure gradient at the outer edge is large enough to push
the dust outward. The densest ring outside the planet's orbit forms where $a$ 
switches sign at $r=1.3$. Particles approach this location in the disk from 
both smaller and larger radii, which leads to a strong dust accumulation.

The close-up in Fig. \ref{fig6} indicates the outer 2:1 mean motion 
resonance, which shows up as a bump in the radial acceleration at $r=1.6$.
Note that $a$ is still negative near the resonance, so the dust keeps moving
inward. However, near the resonance, the dust particles are decelerated, and 
therefore the dust piles up at that location. From Fig. \ref{fig5}, we can 
also see that the density peaks just outside the gap are close to, but not 
exactly located at the position of the 3:2 and the 2:3 resonance, showing that 
the mechanism by which they are formed is different. In contrast to the region 
close to the shallow gas gap, there is no apparent pressure gradient near the 
2:1 resonance to structure the dust disk. Nonetheless, the resonance does show 
up in the acceleration in Fig. \ref{fig6}, and it is therefore the velocity 
field rather than the density of the gas that makes the dust move into the 2:1 
resonance. 

The adopted initial gas surface density is correct for a distance of 13 AU
in the Minimum Mass Solar Nebula (MMSN). For this location of the planet, the
initial dust mass between $r=1.2$ and $r=1.5$ is $5.4$ Earth Masses (\mearth).
After 500 orbits of the planet, the dust mass in this region has grown to 
$10.9$ \mearth, an enhancement of more than a factor of 2 
(see Fig. \ref{fig5}). This may have serious implications for building 
planetesimals (boulders with sizes larger than 10 centimeters) in this region.

We have not included dust diffusion in our models. Turbulent transport of small
dust grains is not yet fully understood, but it can be described by diffusion 
for the smallest particles \citep{2005ApJ...634.1353J}. To see what
effect diffusion has on dust-gap formation, we gave the dust fluid the same 
turbulent diffusion coefficient as the gas, and included the viscous force 
terms in the dust source update. It turned out that dust diffusion affects
the dust density distribution only at the $1~\%$ level, indicating that this is
not an important process. However, the largest particles we consider may be 
subject to vortex trapping in turbulent eddies \citep{2005ApJ...634.1353J}, 
and for these particles, the diffusion approach may not be entirely valid. 

\subsection{Dependence on particle size}
\label{secsize}
We have seen that particles of 1 mm decouple from the gas due to the 
disk perturbation induced by a small planet. As we examine smaller and smaller 
particles, we expect a certain minimum particle size for this decoupling to 
occur because smaller particles couple better to the gas. We can make a simple
estimate for this $s_\mathrm{min}$ by looking at the region in the disk 
near $r=1.1$, where the dust experiences the largest positive acceleration
$a_\mathrm{max}$. According to Fig. \ref{fig6}, $a_\mathrm{max}=0.005$. 
The radial gas velocity in a standard accretion disk is
\begin{equation}
\label{eqgasvel}
v_{r,\mathrm{g}}=-\frac{3 \nu}{2 r}.
\end{equation}
The gas radial velocity in a disk with a planet is similar.
Balancing the outward acceleration with the radial drag force gives an
equilibrium velocity difference between gas and dust of
\begin{equation}
\Delta v_r = a_\mathrm{max} \frac{T_\mathrm{s}}{\Omega_\mathrm{K}}.
\end{equation}
When this velocity difference is larger than the radial gas velocity, 
dust particles can be accelerated outward around $r=1.1$, creating a low
density gap and a high density ring. The minimum stopping time for this 
to occur is
\begin{equation}
\label{eqminTs}
T_\mathrm{s,min} = \frac{3 \Omega_\mathrm{K} \nu}{2 r a_\mathrm{max}}.
\end{equation}
For our disk parameters, this implies a minimum particle size of 
approximately 500 $\mathrm{\mu m}$. Note, however, that although smaller
particles will not be pushed outward, they will travel inward at a different
speed near the edge of the gas dip. As in the case for the 2:1 resonance,
where in spite of the acceleration being negative everywhere a dust ring still
arises, these particles will eventually create a gap, but on a much longer 
time scale.

\begin{figure}
\resizebox{\hsize}{!}{\includegraphics[bb=35 10 285 250]{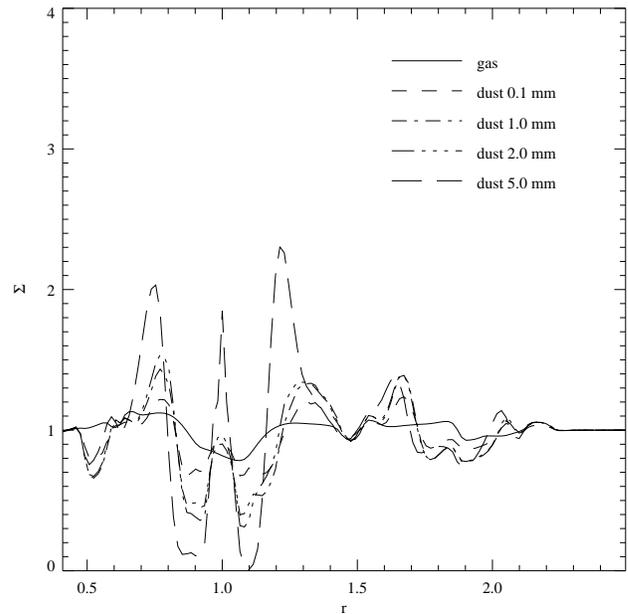}}
\caption{Radial density cut opposite to the planet for four different 
particle sizes after 100 orbits of a $0.1$ \mjups planet. The dust densities 
are multiplied by 100.}
\label{fig7}
\end{figure}

Figure \ref{fig7} shows the radial dust density for four different particle
sizes after 100 orbits. First of all, it is clear that the larger particles
react even more dramatically on the planet. The three bumps at $r=0.7$, 
$r=1.3$, and at corotation all grow very fast with increasing particle size. 
After 400 orbits, the dust-to-gas ratio of 5 mm particles at $r=1.3$ is 
enhanced by more than an order of magnitude. The width of the dust gap does 
not depend on the size of the particles.

The outer disk ($r>1.1$) reaches a quasi-static state after approximately
100 orbits. From then on the only evolution is in the further growth of 
existing features. The inner disk is slowly cleared, however, but on a time 
scale that is set by the unperturbed motion of gas and dust. If we take Eq. 
\ref{eqgasvel} for the velocity of the gas and Eq. \ref{eqvdrift} for the 
relative velocity, we can write the following for the dust velocity:
\begin{equation}
v_r=-\left(\frac{3}{2}\alpha + T_\mathrm{s} \right) h^2 v_\mathrm{K}.
\end{equation}
This gives a time scale for clearing the region inside $r=1$ of approximately
$50000$ orbits.

Looking at the distribution of $0.1$ mm particles in Fig. \ref{fig7}, we see
that even these well-coupled particles start to create a clearing in the 
inner disk. However, since this planet is not strong enough to keep all the
dust outside of $r=1.3$, the inner disk is still fed from outside. Particles 
larger than 500 $\mathrm{\mu m}$ are able to create a fast gap for this 
planet. This agrees with the estimate of Eq. \ref{eqminTs}. Based on Fig.
\ref{fig7}, we estimate that particles larger than $150$ $\mathrm{\mu m}$ will
open up a dust gap within 500 orbits.

\subsection{Dependence on planetary mass}

\begin{figure}
\resizebox{\hsize}{!}{\includegraphics[bb=35 10 285 250]{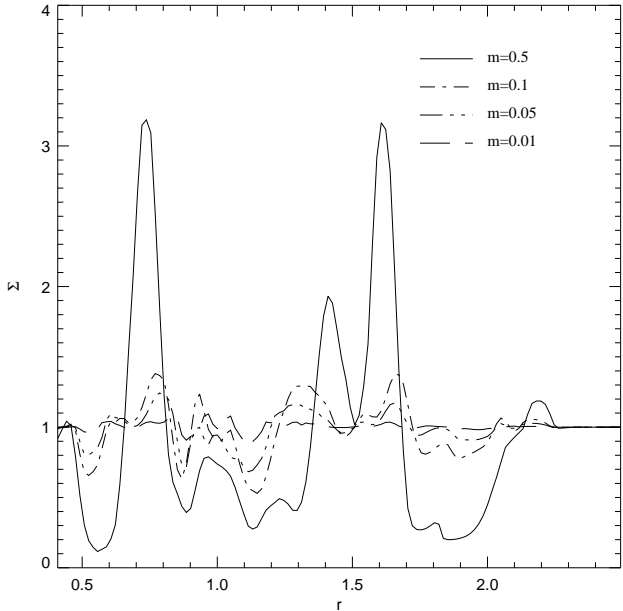}}
\caption{Radial density cut opposite to the planet for four different 
planetary masses after 100 orbits and for a particle size of 1 mm. The dust 
densities are multiplied by 100, and the planetary masses are in units of 
\mjup}
\label{fig8}
\end{figure}

Smaller planets do not dramatically perturb the disk, so we expect that a 
minimum planetary mass for dust-gap opening also exists. In this section, we 
consider only 1 mm particles.

Figure \ref{fig8} shows the radial dust distributions for four different
planetary masses. The smallest planet is not able to produce a density dip
near its orbit, so the gas density is equal to 1 everywhere except in the 
spiral waves. From Fig. \ref{fig8}, we conclude that a planet of $0.05$ 
\mjups is the lowest-mass planet that is able to create a gap, albeit on a 
longer time scale than for more massive planets. 

\begin{figure}
\resizebox{\hsize}{!}{\includegraphics[]{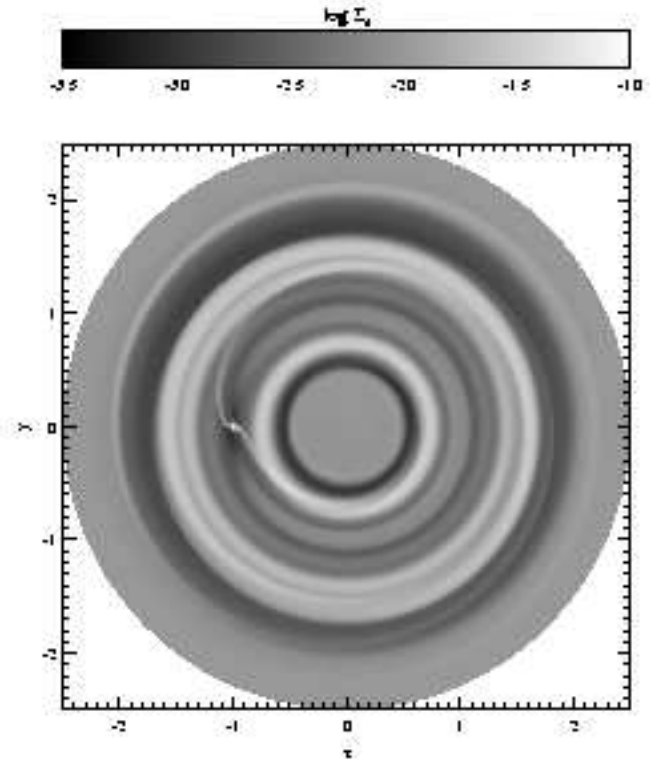}}
\caption{Surface density of 1 mm dust particles after 100 orbits of a 
$0.5$ \mjups planet.}
\label{fig9}
\end{figure}

The highest mass planet in Fig. \ref{fig8} is on the edge of creating a 
gap in the gas disk. As a result of the lower gas density, the stopping time 
for particles near the orbit of the planet is relatively large, which allows 
for the strong evolution  near $r=1$. The time scale for dust-gap 
formation for this planet is about half the time scale for gas-gap formation, 
which is approximately 100 planetary orbits. Also, due to the strong spiral 
waves, the 2:1 mean motion resonance plays a major role; the 
resonance is able to suck material up from both sides, leading to a less
pronounced peak near $r=1.4$ and to an empty region around $r=1.8$. Figure 
\ref{fig9} shows the dust surface density for this planet. Note that the 
resonance-induced gaps at $r=0.55$ and $r=1.8$ are as deep as the gap near the 
planet's orbit. Observationally, it may be difficult to distinguish a 
single-planet system like this from one with multiple planets.

\begin{figure}
\resizebox{\hsize}{!}{\includegraphics[bb=25 10 315 245]{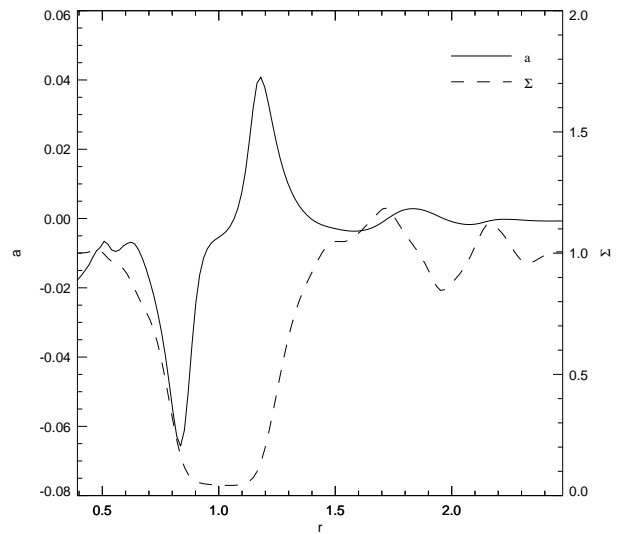}}
\caption{Azimuthal averages of the radial acceleration $a$ (solid line) and 
gas surface density $\Sigma$ (dashed line, shown for $0.4 \le r \le 1.8$)
for a 1 \mjups planet after 400 orbits.}
\label{fig10}
\end{figure}

\begin{figure}
\resizebox{\hsize}{!}{\includegraphics[]{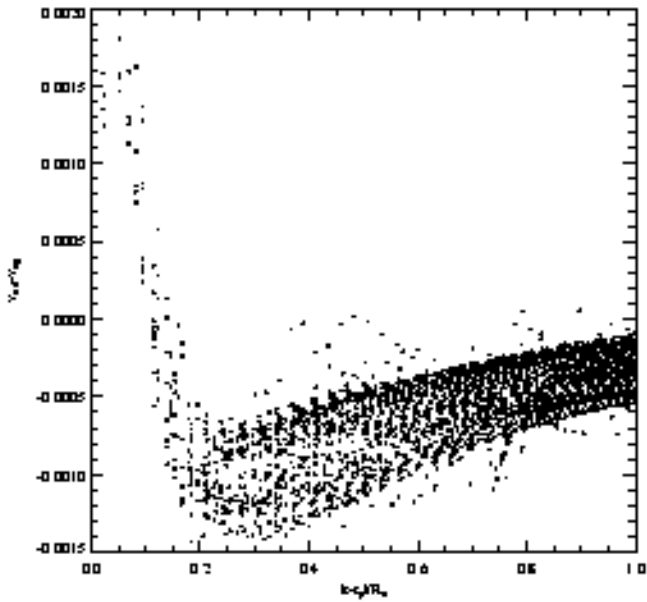}}
\caption{Scatter plot of the relative velocity in the direction of a 0.1 
\mjups planet. The distance to the planet is in units of the Roche lobe 
$\mathrm{R_R}$, and the unit of relative velocity is the Kepler velocity at 
the position of the planet. The size of the dust particles in this simulation 
is 2 mm.}
\label{fig11}
\end{figure}

Because of the conditions discussed in Sec. \ref{secLim}, a $0.5$ \mjups 
planet is the maximum planetary mass we can accurately model with a two-fluid
calculation. However, based on the relation between the radial acceleration
$a$ and the resulting dust distribution (compare Figs. \ref{fig5} and 
\ref{fig6}), we can still make a few remarks concerning higher-mass planets. 

First of all, we expect similar things to happen for a 1 \mjups planet as for a
$0.5$ \mjups planet, only in a more extreme form. Secondly, we noted in Sec. 
\ref{secGlob} that the dust ring just outside the
gap forms where $a$ changes sign from positive to negative. In Fig. 
\ref{fig10}, we show $a$ together with the gas surface density $\Sigma$ for a 
1 \mjups planet.  The gas gap clearly stands out, and we see that $a$ changes 
sign at the outer edge of the gap around $r=1.4$. A dust ring at that location
would coincide with the outer edge of the gas gap. Therefore, a possible dust 
gap would have the same width as the gap in the gas disk.

Also from Fig. \ref{fig10}, we see that $a$ is structured up to
$r \approx 2.5$, which is larger than for the $0.1$ \mjups planet (see Fig. 
\ref{fig6}). For a planet at Neptune's orbit (semi-major axis 30 AU), the 
resonant structure in the dust disk can therefore extend to at least 75 
AU.

\begin{figure*}
\includegraphics[bb=25 10 530 245,width=17cm]{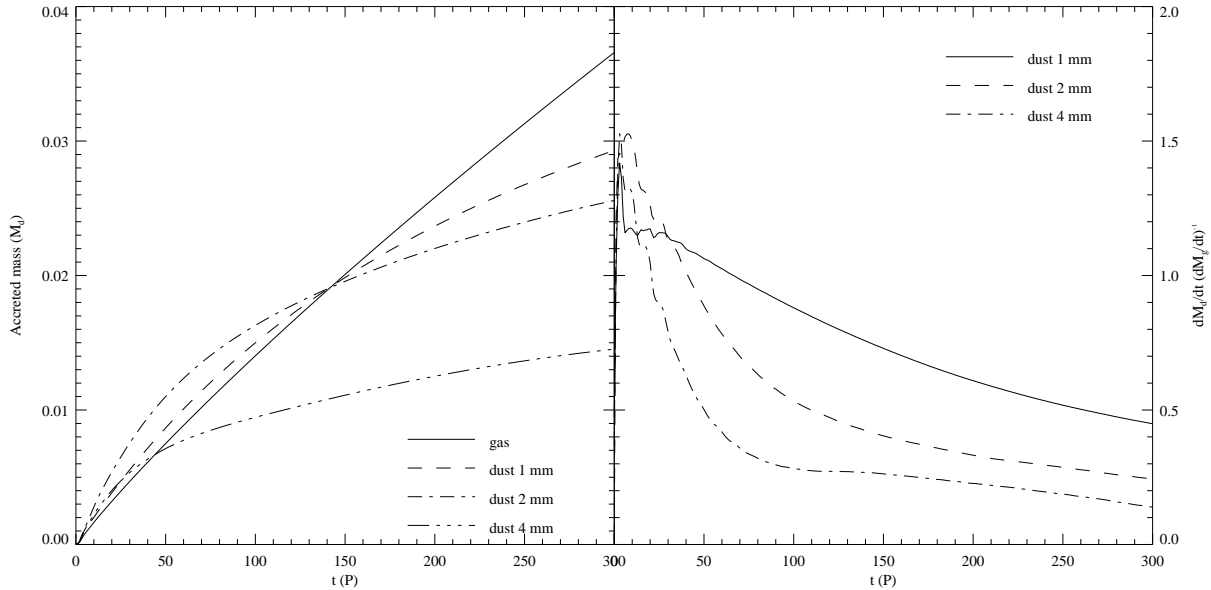}
\caption{Accretion of gas and dust onto a 0.1 \mjups planet. Left panel:
gas and dust mass accreted on the planet. Right panel: relative accretion
rate of dust with respect to the gas. For both panels, the dust mass is
multiplied by 100 to compare it with the gas.}
\label{fig12}
\end{figure*}

\subsection{Flow within the Roche lobe}
The flow of dust deep within the Roche lobe ($\mathrm{R_R}$) of the planet is 
relatively simple. Due to the strong pressure gradient in the direction of the 
planet, the relative velocity of the dust particles is always directed towards 
the planet. In other words, if we take a cylindrical coordinate frame 
$(s,\theta)$ centred on the planet, then $\Delta v_s = v_{s,d}-v_{s,g}$ is 
always negative.

In Fig. \ref{fig11}, we show this relative velocity as a function of distance
to a $0.1$ \mjups planet for 1 mm particles. The first thing to note is that 
the relative velocity changes sign very close to the planet. That is because 
this was a run with gas and dust accretion, and therefore most of the gas 
inside $0.1~ \mathrm{R_R}$ was removed (see \cite{paardekooper03} for the 
accretion procedure), leading to a reversed pressure gradient locally. Outside 
the accretion area, dust moves slowly towards the planet with respect to the
gas. The spread in velocities stems from the fact that the gas density 
distribution around the planet is not symmetric with respect to the planet 
\citep{2002A&A...385..647D}. 

If we take the average relative velocity in the Roche lobe as $4.0~10^{-4}$
in units of the Kepler velocity at the location of the planet, then we get a 
time scale to clear the Roche lobe of dust of less then 
$100$ orbits for particles of 1 mm. This is an important time scale for 
accretion, as we will see in the next section.

\subsection{Gas and dust accretion}
\label{secAcc}
We now turn to the interesting problem of dust accretion. The accretion
of dust particles compared to the gas is important for the final enrichment
of gaseous giant planets. When the dust accretion rate is higher than 0.01 (the
initial dust-to-gas ratio) times the gas accretion, the planet will get 
enriched in solids, and when the dust accretion is lower than $0.01$ the 
gas-dust mixture that is accreted onto the planet is relatively gas-rich. 
In the discussion below, we multiply the dust content of the disk by 100, so 
that gas and dust accretion can be directly compared.

The mass range we can consider is limited because for $M_p < 0.1$ \mjups, 
three-dimensional effects start to play an important role 
\citep{2003ApJ...586..540D}, while planets larger than approximately $0.5$ 
\mjups open up a deep gas gap in the disk, which we cannot treat in the 
fluid approach. Therefore, to eludicate the basic principles that govern dust 
accretion and to provide a starting point for future, more detailed 
simulations, we only consider a $0.1$ \mjups planet.

In the left panel of Fig. \ref{fig12}, we show the accreted mass of gas and 
dust in units of gas and dust disk masses as a function of time. The 
accretion of gas (solid line) was extensively discussed in 
\cite{paardekooper03} and is shown here only for comparison. After 
approximately $200$ orbits, it reaches a constant value of $1.5~10^{-4}$ gas 
disk masses per orbit.

The accretion rates for dust particles of different sizes show a different
behaviour. In the first few orbits, relatively more dust is accreted than 
gas, which means that the dust mass that is accreted is larger than $0.01$
times the gas mass that is accreted. This is the stage where the immediate 
surroundings of the Roche lobe are cleared of dust, due to the strong pressure 
gradients in the planetary atmosphere. Later, the competing process of 
dust-gap formation sets in, and consequently, dust accretion slows down. In the
end, for all particle sizes, relatively more gas is accreted than dust, which 
means that the planet will not be enriched in solids of these particular sizes.

The right panel of Fig. \ref{fig12} shows the ratio of accretion rates for the 
different particle sizes to the gas accretion rate. We see that initially the 
accretion rate of dust is higher than 0.01 times the accretion rate of the gas
or all particle sizes, corresponding to the Roche lobe clearing 
mentioned before. If we look at the accretion rate of 1 mm particles, after 
the first sharp peak, the accretion rate stays almost constant for 30 orbits, 
after which the process of gap formation sets in, and the accretion rate 
slowly declines. After 300 orbits, the accretion rate for the dust is less 
than half the accretion rate of the gas, and it is still declining.

For the larger particles, we see a larger initial accretion rate
because these more weakly coupled dust particles move up the strong pressure 
gradient near the planet more easily. However, for larger particles,
gap formation sets in earlier, as well, and it will eventually win. A particle
size of 2 mm results in the largest accreted mass of dust in this first 
stage of accretion. For 4 mm particles, gap formation sets in before the planet
can accrete a significant amount of surrounding dust. However, in the long
run, more small than large particles will have been accreted, as can be seen 
from the left panel of Fig. \ref{fig12}. 

The planet clears the outer part of the gap of 4 mm particles in approximately
100 orbits (see Fig. \ref{fig7}). At this point, the accretion rate curve 
levels off at a value of $0.2~\dot{M}_{\rm gas}$, but around 150 orbits, it 
starts to decrease again. Accretion around this time comes from the dense
ring at $r=1$, which will slowly diffuse into the inner disk. After it is 
gone, dust accretion will be negligible compared to gas accretion.

\section{Discussion}
\label{secDisc}
Dust gaps opened by low-mass planets have some interesting observational and 
theoretical consequences. In \cite{2004A&A...425L...9P}, it was shown that the 
Atacama Large Millimeter Array (ALMA) will provide the resolution to observe 
these gaps in protoplanetary disks at the distance of the Taurus starforming
region. Observationally, the relatively large particles we consider are only 
important at mm-wavelengths. To create gaps and inner holes in disks 
that can be detected at infra-red wavelengths, more massive planets are needed 
to remove the smallest particles. 

Next, we will explore some more possible consequences for planet formation.
Before we can make any quantitative statements about the importance of the 
limited size range we consider in this paper, we need to determine what 
fraction of the total dust mass these particles represent. If we look at the 
size distribution from Fig. 7 of \cite{2005A&A...434..971D} we see a 
bimodal distribution (dust and planetesimals), with a transition around 
particles of 10 cm. It is interesting to note that for typical mid-plane 
densities, a size of 10 cm gives rise to $T_{\rm s} \approx 1$ (see Eq. 
\ref{eqstopinit}), which is the upper limit for which the fluid approach is 
valid. Boulders larger than 10 cm are not influenced by the gas as much as the 
smaller particles, and therefore will not participate in gap formation due to 
pressure gradients, while particles smaller than $150$ $\rm{\mu m}$ are too 
well-coupled to the gas to form a gap (see Sec. \ref{secsize}). From Fig. 7 of 
\cite{2005A&A...434..971D}, we estimate that half of the total mass of 
particles smaller than 10 cm is in particles larger than $150$ $\rm{\mu m}$. 
Furthermore, about $30~\%$ of the total dust mass is in particles with sizes 
between $150$ $\rm{\mu m}$ and $10$ cm. 

The above estimate implies that the inner disk is denied $30~\%$ of the amount 
of solids it would receive from the outer disk without the presence of the 
planet. If (terrestrial) planets are still being formed in the inner disk, it 
may slow down this process considerably. 

Dust piles up outside the planet's orbit the (see Fig. \ref{fig5}). 
In Sect. \ref{secGlob}, we mentioned that after $500$ orbits of the planet,
the amount of dust had already doubled. Because the planet acts as a barrier
for particles moving inward, more and more dust will stream into the resonance 
from the outer disk. 
It is interesting to note that this is in the region where the Kuiper Belt
resides with respect to Neptune. This large amount of dust may well have 
influenced planetesimal formation in that region.

If an already-formed giant planet is present in the inner disk, 
this planet is also denied $30~\%$ 
of the amount of solids it would be able to accrete without the presence of 
the outer planet. One could imagine that in the present configuration of
our Solar System, with three dust-gap-opening planets outside $5.2$ AU,
the amount of solid material accreted by Jupiter would be dramatically small,
even without the mechanism outlined in Sect. \ref{secAcc}.

Depending on the dust size distribution in the disk, the low relative
accretion rates for mm-sized particles found in Sect. \ref{secAcc} will have 
a significant effect on the
final composition of the planet. Our results show that the accretion rates for 
these particles are, on the long run, more than an order of magnitude lower 
than the gas accretion rate. This will significantly affect the final 
enrichment in solids for giant planets, even more so if a dust-gap-opening 
planet exists in the outer disk.

In other words, if, as is standard in planet-formation models,
one considers gas accretion and the accretion of planetesimals separately, 
the gas that is accreted is 
relatively {\it poor} in solids (by a factor of 2) compared to the solar value.
This process works against the usual enrichment scenarios for the giant 
planets \citep{2001ApJ...550L.227G, 2001ApJ...559L.183G, 1999Natur.402..269O}. 
However, as was mentioned before, we cannot make firm statements about planets 
of higher mass. More definitive conclusions about dust accretion onto giant 
planets require further study.

Dust growth may also play an important role in structuring the disk. 
When the growth time scale $\tau_g$ is long compared to the dust-gap-formation 
time scale, what effectively happens is that particles will be removed from the
dust gap as soon as they become large enough. This means that the gap will 
also be cleared of the smallest particles on a time scale $\tau_g$. 
However, the process of dust growth in disks is not understood well enough 
to make any quantitative statements about it.

It is important to realize that the dust-gap-formation process basically only
needs a shallow gas dip formed by an embedded planet. As long as more 
complicated disk models including magnetic fields and radiative transfer do
not change this gas structure, the basic picture outlined in this paper will
not change. Three-dimensional 
effects may be important, however, because gas and dust behave differently
in the vertical direction: dust particles will settle to the midplane, while
the gas forms a pressure-supported structure. Angular momentum transport, 
and therefore also gap formation, is reduced in three-dimensional models,
compared to vertically integrated models, but because of settling effects, the 
dust will mainly interact with gas near the midplane where the planetary
potential is strongest. It is not clear how these competing processes will
affect our results; three-dimensional, multi-fluid simulations are needed
to answer that question.

\section{Summary and conclusions}
\label{seccon}
In this paper, we have elaborated on the findings of 
\cite{2004A&A...425L...9P},
who showed that intermediate mass planets are able to open dust gaps in gas 
disks. We have shown how the formation of the gap depends on particle size
and planetary mass, indicating that particles larger than 150 $\rm{\mu m}$ and
planets larger than $0.05$ \mjups are needed to open a gap in a typical 
protoplanetary disk. Planets larger than $0.5$ \mjups clear a gas gap
around their orbit, increasing the stopping time $T_\mathrm{s}$ locally by
2 orders of magnitude, leaving the dust inside the gap largely decoupled from 
the gas. If the planet clears a gas gap, the dust gap will be of equal width.

Within the Roche lobe of the planet, the pressure gradient is such that 
solids will quickly move to the centre of the planetary atmosphere, allowing 
for a very high dust accretion rate. However, this enhancement is more than
counterbalanced by the local depletion of dust due to the formation of 
the dust gap. In the end, much more gas can be accreted compared to solids. 
This may have serious consequences for the enrichment of giant planets. 

\begin{acknowledgements}
S.P. and G.M. acknowledge financial support from the European Research 
Training Network ``The Origin of Planetary Systems'' (PLANETS, contract number 
HPRN-CT-2002-00308) at Leiden Observatory. G.M.'s work in Leiden is made 
possible through support from the Royal Netherlands Academy of Arts and 
Sciences.
\end{acknowledgements}

\bibliographystyle{aa} 
\bibliography{4449.bib}

\end{document}